# Intrinsic signal optoretinography of dark adaptation kinetics

Tae-Hoon Kim[1], Jie Ding[1], and Xincheng Yao[1,2*]

[1]Department of Biomedical Engineering, University of Illinois at Chicago, Chicago, IL, 60607, USA

[2]Department of Ophthalmology and Visual Sciences, University of Illinois at Chicago, Chicago, IL, 60612, USA

*Corresponding author: xcy@uic.edu

**Abstract:** Delayed dark adaptation due to impaired rod photoreceptor homeostasis has been reported as the earliest symptom of eye diseases such as age-related macular degeneration, diabetic retinopathy, and retinitis pigmentosa. Objective measurement of dark adaptation can facilitate early diagnosis to enable prompt intervention to prevent vision losses. However, there is a lack of noninvasive methods capable of spatiotemporal monitoring of photoreceptor changes during dark adaptation. Here we demonstrate functional optical coherence tomography (OCT) for in vivo intrinsic signal optoretinography (ORG) of dark adaptation kinetics in the C57BL/6J mouse retina. Functional OCT revealed a shortening of the outer retina, a morphological change in the cone and rod photoreceptor interdigitation zone, and a reduction in intrinsic signal amplitude at the photoreceptor inner segment ellipsoid. A strong positive correlation between morphophysiological activities was also confirmed. Functional OCT of dark adaptation kinetics promises a method for rapid ORG assessment of physiological integrity of retinal photoreceptors.

**Introduction**

Dark adaptation refers to a systematic recovery of visual sensitivity in the dark following exposure to bright lights[1]. During dark adaptation, multiple processes simultaneously occur to maintain retinal homeostasis, including photoreceptor repolarization[2], photopigment regeneration[2], blood flow modulation[3], a redistribution of photoreceptor signaling proteins[4] and interphotoreceptor matrix proteins[5], and a reversal of metabolic energy flow[6]. Visual perception in night vision is rod-mediated; thus, delayed dark adaptation due to impaired rod photoreceptor homeostasis is often the earliest symptom of various retinal diseases, such as age-related macular degeneration (AMD)[7], diabetic retinopathy (DR)[8], and retinitis pigmentosa (RP)[9], the leading causes of irreversible vision loss. Since early detection is essential to enable prompt treatment to prevent vision loss[10, 11, 12, 13], the measurement of dark adaptation kinetics has been proposed as a potential solution for screening early-stage retinal diseases.

Conventional dark adaptation measurement is psychophysical, i.e., perimetry, or electrophysiological, i.e., electroretinogram (ERG)[14, 15]. However, the perimetry test is subjective in nature, primarily influenced by the reliability of patient performance[16], and reveals low threshold sensitivities so that subtle abnormality correlated with early diseases might not be reliably detected[17]. While ERG can provide objective and quantitative information on retinal dysfunction, it represents the integral activity of whole retinal neurons. Accordingly, the pathological condition of any retinal neurons may affect the electrical loop of ERG recording. Considering the coherent interactions between retinal cells, it appears necessary to quantitatively examine anatomical arrangement simultaneously as well as functional activity.

Recent developments of intrinsic signal optoretinography (ORG) promise a new methodology for the noninvasive objective assessment of dark adaptation. ORG generally refers to functional optical imaging of the retina. Functional optical coherence tomography (OCT) promises a depth-resolved imaging modality to enable intrinsic signal ORG of the retina[18, 19, 20, 21]. Functional OCT has revealed intrinsic optical signal (IOS) distortions correlated with pathophysiological conditions in diseased animal model[22]. Previous OCT studies of the dark-adapted mouse retina revealed outer retinal shortening along with a reduction in the width of a hypo-reflective band between the photoreceptor outer segments (OS) and RPE[23, 24, 25]. However, dynamic monitoring of transient morphophysiological activities, which promises a biomarker for rapid assessment of functional integrity of dark adaptation in the retina, remains to be explored.

Here we report intrinsic signal ORG assessment of dark adaptation kinetics in the mouse retina in vivo. The light-adapted retina was initially measured as a baseline, followed by subsequent recording under dark conditions.

Dark-induced morphophysiological changes in the outer retina were clearly observed in OCT imaging. High-speed monitoring further identified a strong correlation between morphological and physiological changes in the outer retina. This study demonstrates the potential of intrinsic signal ORG assessment of dark adaptation kinetics in the living retina.

**Results**

**Imaging features of light- and dark-adapted mouse retinas:** Adult C57BL/6J mice were used in this study. Since rods constitute ~97% of mouse retinal photoreceptors, the dark-induced retinal activity measured in this study was mainly governed by the canonical visual cycle[1]. The experiment was performed during the early afternoon when the mouse retina was light-adapted by constant ambient light conditions. Total seven retinal OCT volumes were sequentially acquired at the same region on the dorsal quadrant per each mouse; 1st OCT for light-adapted baseline measurement and 2nd to 7th OCTs for dark adaptation measurement, acquired at every 5 min interval for 30 min. To first validate the adaptation effects on the retina, a direct comparison between the light-adapted (1st OCT) and dark-adapted (7th OCT) retinas was performed. Figure 1A shows representative OCT images from different mouse retinas. We noted three distinct features in the retina under different light conditions (Fig. 1B). First, it was readily recognizable that dark adaptation caused a reduction in outer retinal thickness. Second, in the light-adapted retina, the 3rd band of the outer retina was distinguished, which was absent in the dark-adapted retina, and a hypo-reflective band between the 3rd and 4th band was only observed in the light-adapted retina. Third, OCT intensity in the 2nd outer retinal band markedly decreased in the dark-adapted retina. These results demonstrate that dark adaptation induces morphophysiological changes in the retina.

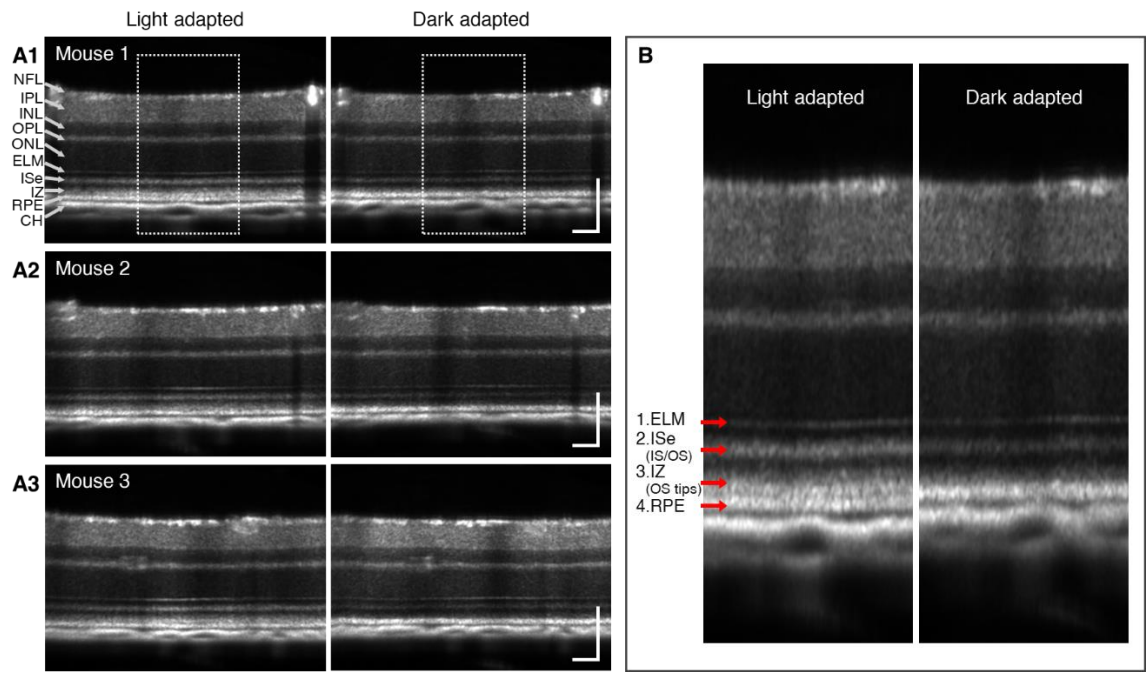

**Figure 1.** Comparison of light- and dark-adapted retinas. (A) OCT images of light- and dark-adapted retinas from 3 different mice (A1-A3). (B) Enlarged view of the white rectangle regions in (A1). Red arrows indicate OCT hyper-reflective bands in the outer retina. NFL: nerve fiber layer; IPL: inner plexiform layer; INL: inner nuclear layer; OPL: outer plexiform layer; ONL: outer nuclear layer; ELM: external limiting membrane; ISe: inner segment ellipsoid; IZ: interdigitation zone; RPE: retinal pigment epithelium; CH: choroid; IS: inner segment; OS: outer segment. Scale bars: 100 µm.

**Quantitative analysis of morphophysiological changes in the retina during dark adaptation:** Figure 2A shows repeated measures of OCT during 30 min dark adaptation. To quantify retinal changes, average OCT A-lines were prepared and used for retinal thickness comparison at different time points. We found that the inner retina showed little change (Fig. 2B), while the outer retina revealed a significant reduction in thickness during dark adaptation (Fig. 2C). A repeated-measures analysis of variance (ANOVA) determined that mean outer retinal thickness differed significantly between different time points ($p < 0.001$). Post-hoc tests using the Bonferroni correction revealed that even 5 min dark adaptation resulted in a significant reduction in outer retinal thickness (Light: 106.8 ± 2 µm; 5 min dark: 105.1 ± 1.9 µm; $p < 0.001$). However, dark adaptation had not significantly altered inner retinal thickness (Light: 79.42 ± 1.7 µm; 30 min dark: 78.65 ± 2 µm; $p = 0.15$). Next, the outer retina was segmented into two regions and analyzed (Fig. 2D, E). A slight decline was observed in the outer nuclear layer (ONL) thickness (Fig. 2D). In contrast, the external limiting membrane (ELM)-RPE complex revealed a rapid decline in thickness right after the light went off. A repeated-measures ANOVA determined that mean ONL ($p < 0.001$) and ELM-RPE ($p < 0.001$) thickness differed significantly between the time points, and post-hoc tests using the Bonferroni correction revealed that statistically significant changes were first observed after 20 min dark adaptation in the ONL (Light: 58.08 ± 1.6 µm; 20 min dark: 56.76 ± 1.5 µm; p < 0.01) and 5 min dark adaptation in the ELM-RPE complex (Light: 48.73 ± 1.6 µm; 5 min dark: 47.08 ± 1.4 µm; p < 0.001). These results are in line with previous observations[25, 26] and demonstrate that ELM-RPE complex predominantly affects retinal thinning, accounting for ~70% reduction of the outer retina (ELM-RPE displacement after 30 min dark: -4.07 µm; total outer retinal displacement after 30 min dark: -5.83 µm).

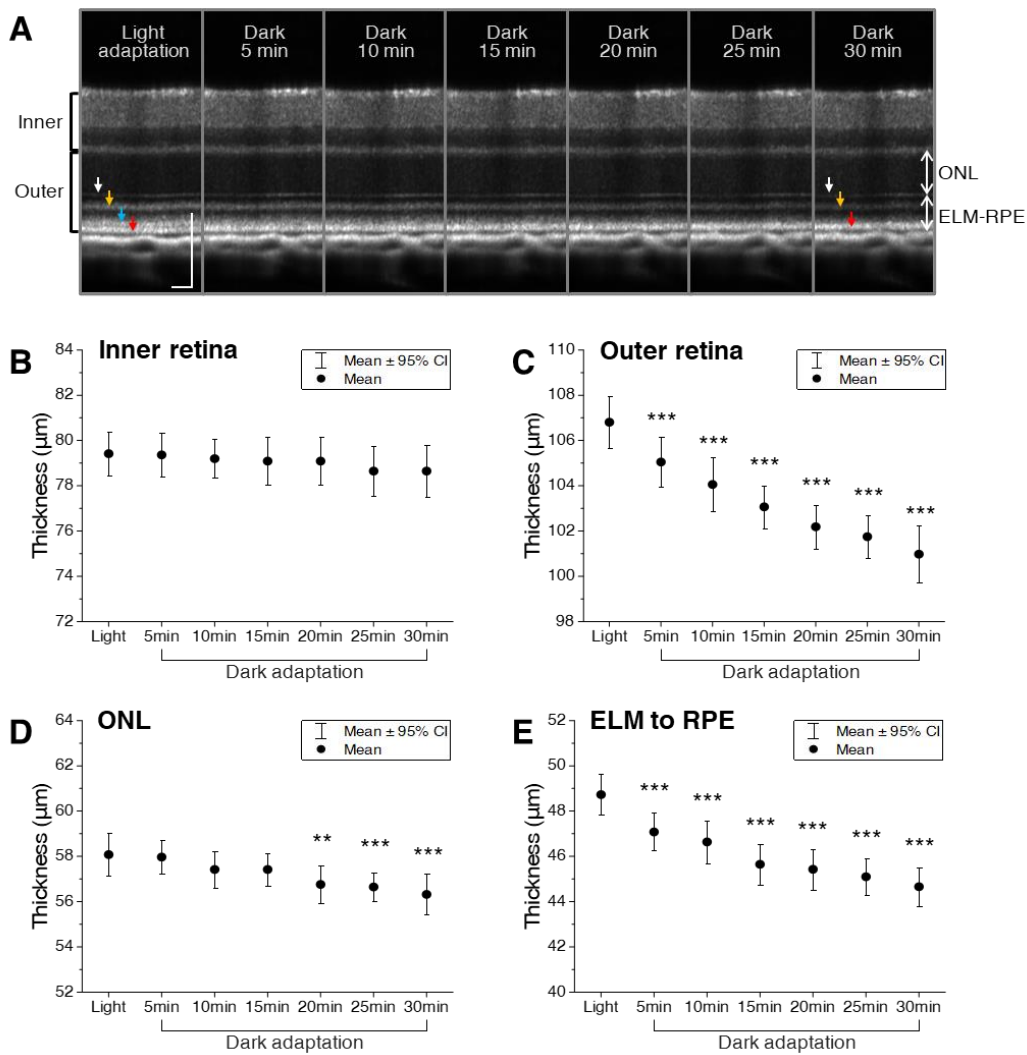

**Figure 2.** Quantitative measurement of retinal thickness changes during dark adaptation. (A) A sequence of OCT images was obtained every 5 min up to 30 min during dark adaptation. Arrows to indicate outer retinal bands: 1st ELM band (white), 2nd ISe and IS/OS band (yellow), 3rd IZ and OS tip band (blue), 4th RPE band (red). ONL: outer nuclear layer; ELM: external limiting membrane; RPE: retinal pigment epithelium; ISe: inner segment ellipsoid; IS: inner segment; OS: outer segment. Thickness measurement of (B) the inner retina, (C) outer retina, (D) ONL, and (E) ELM-RPE complex during dark adaptation. Data are represented as mean ± C.I. Significant differences between time points were tested using a one-way repeated measures ANOVA

with Bonferroni's post-hoc tests. N = 14. *P < 0.05, **P < 0.01, ***P < 0.001 compared to the light-adapted retina. Scale bars: 100 μm.

OCT signal is highly sensitive to changes of intrinsic optical properties in excitable cells, such as light scattering, polarization, and absorption fluctuations[27, 28]. As the OCT intensity reduction was apparent in the 2nd outer retinal band (Fig. 1B), we further analyzed relative intensity changes in different hyper-reflective bands, including inner plexiform layer (IPL), outer plexiform layer (OPL), inner segment ellipsoid (ISe), and RPE. The average OCT A-lines were used for the relative intensity measurement after intensity normalization based on the ONL reflectance. We found that the ISe intensity had significantly reduced after 5 min dark adaptation (p < 0.001). Relative to the baseline measurement, the IPL, OPL, and RPE also showed a significant intensity reduction but at the later phase of dark adaptation compared to the ISe change (post-hoc tests using the Bonferroni correction: IPL at 20 min (p < 0.01); OPL at 15 min (p < 0.01); RPE at 20 min (p < 0.05)). Collectively, dark adaptation induces significant IOS changes in OCT, which may reflect physiological activity in the retinal cells.

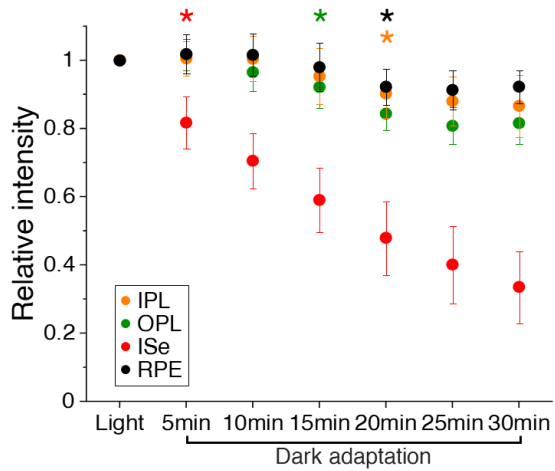

**Figure 3.** Relative OCT intensity measurement of hyper-reflective bands in the retina. Data are represented as mean ± C.I. Significant differences between time points were tested using a one-way repeated measures ANOVA with Bonferroni's post-hoc tests. N = 14. Asterisks indicate the earliest time point showing a statistically significant difference (*P < 0.05) in each band compared to the baseline light-adapted measurement.

**High-speed monitoring of a light-dark transitional period:** Notable retinal changes were observed even after 5 min dark adaptation (Fig. 2 and 3), likely suggesting that OCT may detect early changes of the retina during a short transitional period to darkness. High-speed B-scan recordings were thus performed with 62.5 ms temporal resolution at a fixed retinal plane. In a total 5 min recording, the light-adapted retina as a baseline was imaged for the first 30 sec, and the retina was continuously being imaged for 4.5 min in darkness. The lights were off at 30 sec. Figure 4A exhibits a motion scan (M-scan) illustrating spatiotemporal relations, where each column represents an average A-scan, and individual A-scans were vertically aligned based on the ELM's peak location. Figure 4B1 shows the vertically enlarged view of the outer retina in Fig. 4A, and Fig. 4B2 is the same enlarged image with a color-lookup table to highlight the intensity variations. In line with the previous results (Fig. 1 and 2), it was observed that the 3rd outer retinal band became elusive over time (white arrowheads in Fig. 4B2), and the boundary between the RPE and choroid shifted toward the inner retinal side (green arrows in Fig. 4B2), indicating the outer retina became short.

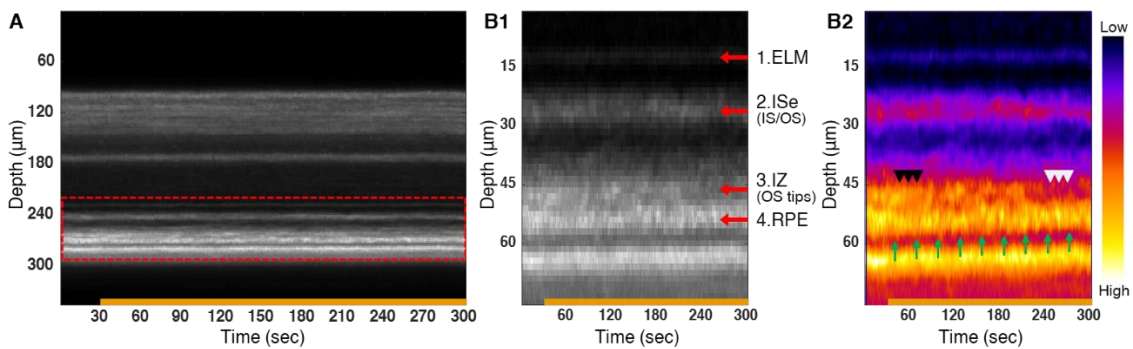

**Figure 4.** High-speed OCT recording during the light-dark transition. The lights were off at 30 sec. (A) Spatiotemporal mapping of OCT images. Each column represents a 1-s recording. (B1) Enlarged view of the red box in (A). (B2) Enlarged view of the red box in (A) with a color-lookup-table "fire" provided in ImageJ software. The range of the lookup table is shown in the bar to the right of the image. Black and white arrowheads point out the 3rd band change during dark adaptation. Green arrows indicate basal RPE position. ELM: external limiting membrane; ISe: inner segment ellipsoid; IS: inner segment; OS: outer segment; IZ: interdigitation zone; RPE: retinal pigment epithelium; CH: Choroid.

Further detailed analysis was made by color-coded intensity profiles (Fig. 5A). Different colors represent different time points; dark red and blue indicate the early and late phases of recording, respectively. From the results, we found three imaging features. First, ISe intensity reduction rapidly occurred after lights off (Fig. 5B). Second, as seen in Fig. 4B2, outer retinal bands, including the RPE and choroid, were being shifted toward the ELM side (Fig. 5C). These observations were consistent with the results presented in Fig. 2 and 3. Third, the color-coded intensity profiles delineate that the 3rd band was being faded over time (Fig. 5D). Intriguingly, we observed not only a peak at the OS tip position but also another peak in the midway of the OS (black arrows in Fig. 5A and D), and both peaks tended to gradually decrease in intensity right after lights off. In addition, the presence of linearity between retinal responses and time courses of dark adaptation was confirmed by Pearson correlation tests (Fig. 5E, F). Results showed that dark adaptation time was negatively correlated with ISe intensity (r = -0.933) and ELM-RPE displacement (r = -0.982). We also found a strong positive relationship between ISe intensity and ELM-RPE displacement (r = +0.944) (Fig. 5G). Collectively, these data suggest that retinal changes gradually occur in a linear manner upon darkness, and high-speed spatiotemporal mapping enables the measurement of prompt retinal activities due to dark adaptation.

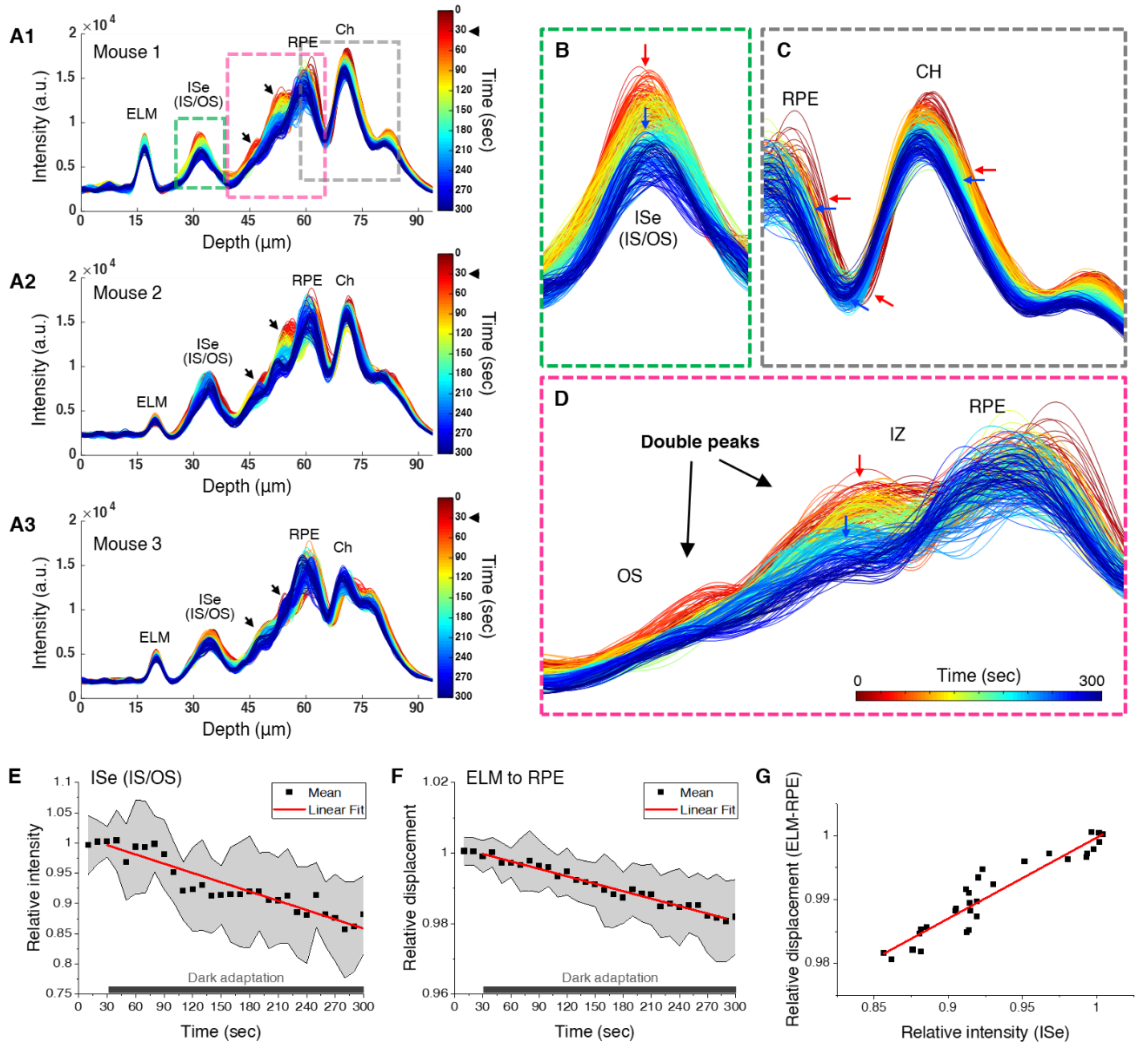

**Figure 5.** Quantitative analysis of high-speed OCT recording during transition to darkness. (A1-3) Color-coded OCT intensity profiles of the outer retina from 3 different mice. Red and blue colors indicate a temporal gradient of a 30-min dark adaptation period. The lights were off at 30 sec. Two black arrows indicate double peaks found along the photoreceptor OS. (B) Enlarged view of the ISe band, green box in (A1). (C) Enlarged view of the RPE and CH, grey box in (A1). (D) Enlarged view of the OS, IZ, and RPE, pink box in (A1). (E) Relative intensity changes and (F) ELM-RPE displacement during dark adaptation. Black dots

represent the mean values, and the accompanied grey area represents standard deviation. (G) Correlation test between ISe intensity and ELM-RPE displacement. Black dots represent the mean values, and a red line is the linear fitting line. ELM: external limiting membrane; RPE: retinal pigment epithelium; ISe: inner segment ellipsoid; IS: inner segment; OS: outer segment; IZ: interdigitation zone; CH: Choroid.

**Double OS peaks may represent cone and rod tips:** An interesting observation in this study was a moderate-reflective peak above the conventional hyper-reflective 3rd band (Fig. 5A, D), suggesting the presence of an additional 3rd outer retinal band. Since this additional peak was consistently observed in 13 out of 14 mice, it was less likely an artifact. The first 3rd band (i.e., the ISe side) was relatively gentle compared to the second 3rd band (i.e., the RPE side), and both peaks gradually decreased in intensity during dark adaptation (Fig. 5A, D). Considering the cone OS is shorter in length and lower in occupancy than the rod OS in the mouse retina, the two layered configuration could be associated with the cone and rod OSs. To verify this notion, a peak-to-peak distance was measured from the ISe to each 3rd band peak, and we found that the distances were 13.4 ± 1.8 µm for the first peak and 20.9 ± 2.2 µm for the second peak (Fig. 6A), well matching the cone and rod OS length in the mouse retina[29]. A representative M-scan image by averaging 14 different M-scans was also constructed to reveal a layered intensity distribution (Fig. 6B1). The distribution was not a continuous gradient along the depth direction; it was rather a discrete staircase distribution (Fig. 6B2). This pattern became more obvious when OCT intensity was depicted with a color-lookup-table (Fig. 6C). The blue-colored OS region represents the proximal OS without an ensheathment by RPE apical processes; while the green and red regions may correspond to the distal OS region with the ensheathment of the cone and rod OS tips, respectively. This ensheathing configuration has been also named as the contact cylinder[30]. Collectively, our results demonstrate that 3rd outer retinal bands in the mouse retina can be divided in two parts; the first moderate-reflective band and second hyper-reflective band can be associated with cone OS tips and rod OS tips, respectively. This configuration is quite similar to an axial positioning of cone and rod OS tips in the perifoveal outer retina in humans[31].

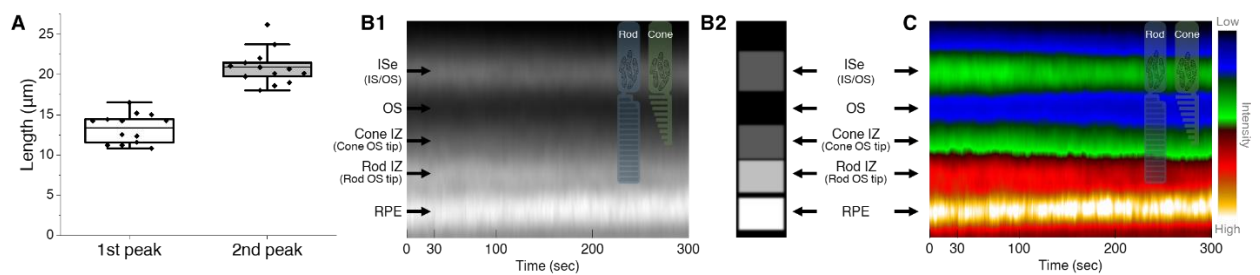

**Figure 6.** (A) Peak-to-peak length measurement from the ISe to double peaks along the photoreceptor OS (N = 13). (B1) An average M-scan image at the outer retinal region. (B2) A simplified intensity distribution model based on (B1). (C) OCT intensity in (B1) is depicted with a color-lookup-table "thal" provided in ImageJ software. The intensity range of the lookup table is shown in the bar to the right of the image. ISe: inner segment ellipsoid; IS: inner segment; OS: outer segment; IZ: interdigitation zone; RPE: retinal pigment epithelium.

**Discussion**

In this study, we demonstrated ORG measurement of dark adaptation in the mouse retina. Functional OCT as an ORG modality was sequentially collected before and after lights off to investigate retinal status under different light conditions. We found that retinal activity during dark adaptation was reflected by morphological (i.e., retinal thickness and OCT bands integrity) and physiological (i.e., intrinsic optical signals) changes, mainly in the outer retina. These responses were manifested as a function of the adaptation time and readily detectable upon darkness. We assumed that coherent interactions between the photoreceptors and RPE to be adapted in the absence of light may be directly or indirectly reflected in OCT imaging.

The altered reflectance observed in the ISe band can support the notion that metabolic activity of the photoreceptors can be associated with IOS changes in the ISe[32, 33]. The photoreceptors display the highest rate of oxidative metabolism in the body and regulate a flow of metabolic energy differentially in light and dark conditions. Energy demands and oxygen consumption are more significant in darkness than in light due to increased adenosine triphosphate (ATP) consumption to maintain the photoreceptor dark current[6, 34]. This enhanced metabolism is mainly managed in the ISe, consisting of densely packed mitochondria. Due to complex internal geometry, mitochondria are the prominent light scatters in OCT imaging, and its geometrical configuration is actively modified based on metabolic demand through the fission and fusion process[35]. It is known that different mitochondrial configurations can alter the scattering property of light[36, 37], which may ultimately change the IOS property of the ISe. In line with this fact, fragmented mitochondria were frequently found in resting cells, while mitochondria became more fused when they are forced to rely on oxidative phosphorylation[38]. A recent study also demonstrated that mitochondria in the cone IS in hibernating ground squirrels were individually smaller and distorted, but mitochondria in active ground squirrels appeared elongated and well-organized. It was hypothesized that the elongated, parallel organization of mitochondria might enhance light delivery to the OS by reducing the number of refractive interfaces[39]. Taken together, it can be plausibly postulated that mitochondria in the ISe may undergo a fusion process during dark adaptation, forming tightly packed and elongated mitochondria to meet the high energy demand, resulting in low reflectance in the ISe band.

We also observed morphological changes of the retina during dark adaptation. The mechanism of outer retinal thinning is rather well understood by RPE-mediated water removal from the ELM-RPE region. As discussed, the transition from light to dark is accompanied by an increase in photoreceptor metabolism, resulting in increased oxygen consumption in the retina[40]. The increase in oxygen consumption leads to a proportionate increase in $CO_2$ and wastewater production into the outer retina, which can acidify the outer retina and ultimately upregulate water removal co-transporters in the RPE. This rapid removal of the acidified water has been linked to significant thinning of the ELM-RPE region[41, 42]. In fact, empirically derived acidic conditions by medicine and hypoxia were found to reduce the outer retinal volume directly[43, 44]. Consistent with our results, Li et al. observed a significant reduction of outer retina thickness in the dark-adapted retina[23]. They also reported that light-dependent volume changes in the outer retina varied with the stage of retinal degeneration in retinal degeneration 10 (rd10) mouse model[26]. Berkowitz et al. found strain-specific changes of the outer retina in different light conditions. A light-driven expansion of the outer retina was more prominent in C57BL/6 mice than 129S6/SvEvTac mice[24]. Gao et al. further demonstrated that dark adaptation significantly reduced the magnitude and width of a hypo-reflective band between the photoreceptor OS and RPE[25]. In addition, Guziewicz et al. showed that the hypo-reflective band became more profound during the transition from dark to light in the canine BEST1 disease model, manifesting retinal abnormalities at the photoreceptor-RPE interface associated with defects in the RPE microvilli ensheathment[45]. Taken together, dynamic ORG measurement of dark adaptation can serve as a functional indicator for RPE-mediated water transport in the outer retina.

However, the water transport could not explain the altered integrity of the 3rd outer retinal bands. We instead hypothesize that interdigitating properties between the OS tip and RPE apical processes may be changed. In this study, we found double peaks along with the OS, and these peaks became enervated during the dark adaptation period. Considering the length from the ELM to each peak (Fig. 6A), the first and second peak most likely correspond to the cone and rod OS tip region, respectively. This region, called the interdigitation zone (IZ), is where the RPE's long apical microvilli interdigitate with the photoreceptor OSs, supporting an adhesion between the retina and RPE. Microvilli are capable of active contraction while interdigitating with the OS[45]. It was demonstrated that the force required to detach the retina from the RPE was 20% greater in the light-adapted retina than in the dark-adapted retina[46]. Also, retinal adhesion in wild-type mice increased by 58% after light adaptation[47]. To support the firm retinal attachment, tight ensheathment might be necessary for light adaptation, presumably manifesting high reflectance in the 3rd band. Conversely, the alleviated tension in darkness along with outer retinal thinning may loosen this interconnection, presumably manifesting low reflectance in the 3rd band. It should be noted that ensheathment of the OS by microvilli is not the only mechanism of retinal adhesion; interphotoreceptor matrix proteins simultaneously contribute to lowering retinal adhesive force in darkness[5, 48].

Another potential reason for the 3rd band attenuation can be attributed to the disc shedding event in which tiny packets of old discs at the distal end of the OS are pruned and digested by the RPE. While disc renewal is the continual assembling process of new discs at the proximal end of the OS, disc shedding is a discrete event influenced by circadian rhythm and light/dark conditions[49]. Disc shedding process is dictated by several distinct steps, including recognition, binding, engulfment, internalization, and digestion[50, 51]. Each step can virtually affect the morphology of the distal end of the OS, i.e., the OS tip. In light, the terminal disc packet might be engulfed by the apical processes of the RPE, while the disc packets may gradually descend upon darkness for internalization. A study reported that during internalization, the disc packets could be rotated as they descend[52], which can alter light reflectance to be less scattered. Similarly, Kocaoglu et al. observed an optical signature of disc shedding, characterized by an abrupt loss in the cone OS tip reflection[53]. It is worth noting that light modulated melanosome translocation can also be a contributing factor in the 3rd band signal[54]. In fact, the 3rd band is the most controversial band among researchers, and multiple interpretations for this region have been proposed[55, 56, 57, 58]. Thus, the exact mechanism remains to be extensively investigated.

In conclusion, we demonstrated the feasibility of in vivo ORG assessment of dark adaptation to characterize dark-induced morphophysiological changes in the mouse retina. Compared to the conventional perimetry and ERG measurements, functional OCT enables intrinsic signal ORG dark adaptation kinetics in a quantitative way with layer-specificity. Moreover, the rapid testing protocol of dark adaptation kinetics would be suitable for both lab research and clinical application because of low subject burden. Although the complex correlations of homeostatic mechanisms and imaging features remain to be elucidated, the intrinsic signal ORG promises a great potential to enable noninvasive, objective measurement of dark adaptation kinetics, which can aid in the diagnosis of early-stage AMD, DR, and RP.

## Methods

### Animal preparation

Two-month-old C57BL/6J mice of either sex (Jackson Laboratory, Bar Harbor, Maine, USA) were used in this study. The mice were kept in regular animal housing under a 14:10 hour light/dark cycle. All animal experiments were approved by the local animal care and biosafety office and performed following the protocols approved by the Animal Care Committee (ACC) at the University of Illinois at Chicago. This study followed the Association for Research in Vision and Ophthalmology Statement for the Use of Animals in Ophthalmic and Vision Research, and we confirm that our work accords with the ARRIVE guidelines.

**Imaging system**

A custom-designed spectral-domain OCT system was used in this study. Technical details of the system are reported in our previous studies[59, 60]. Briefly, a near-infrared super-luminescent diode (λ = 810 nm; Δλ = 100 nm; D-810-HP, Superlum, Carrigtwohill, County Cork, Ireland) was used as the OCT light source. A line CCD camera with 2048 pixels (AViiVA EM4; e2v Technologies, Chelmsford, UK) was used in the custom-built OCT spectrometer. The axial and lateral resolutions of the system were theoretically estimated at 2.9 and 11 µm, respectively. The incident light power on the cornea was 1 mW.

**Experimental procedure**

Imaging was performed in a laboratory room at ~400 lux. Stray light from electronic equipment was masked, and the imaging stage was shrouded with blackout fabric. Experiments were regularly performed between 12 pm to 3 pm. At the time of the experiment, the mouse retina was light-adapted due to constant ambient light conditions. The mouse was first anesthetized by intraperitoneal injection of a mixture of 100 mg/kg ketamine and 5 mg/kg xylazine. A drop of 1% tropicamide ophthalmic solution (Akorn, Lake Forest, IL) was applied to the imaging eye for pupil dilation, and a cover glass (12-545-80; Microscope cover glass, Fisherbrand, Waltham, MA) with a drop of eye gel (GenTeal, Novartis, Basel, Switzerland) was placed on the imaging eye. After the mouse was fully anesthetized, the head was fixed by a bite bar and ear bars in the animal holder. A heating pad was placed around the animal holder to keep the mouse warm.

**Image acquisition**

Volumetric raster scans were performed for OCT imaging. Four repeated B-scans at each slow-scan position were collected; thus, each OCT volume consisted of 4 × 600 × 600 A-scans over 1.2 x 1.2 mm area. Seven OCT volumes were sequentially acquired with 5 min intervals at the same dorsal quadrant. Figure 7 illustrates the flow of image acquisition. The first volume was acquired in light condition, and the other six volumes were acquired in dark condition. High-speed recording to capture the initial moment of dark adaptation was conducted right after the first volume recording. The high-speed recording captured 4800 B-scans at the same retinal plane with 16 frames per second rate for a complete 5-minute recording. While recording, the room light was turned off at 30 sec with the guidance of a LED blink synchronized with the OCT image acquisition system. Data acquisition and real-time imaging preview were made by a custom-built LabVIEW program (LabVIEW 2013, National Instruments, Austin, TX).

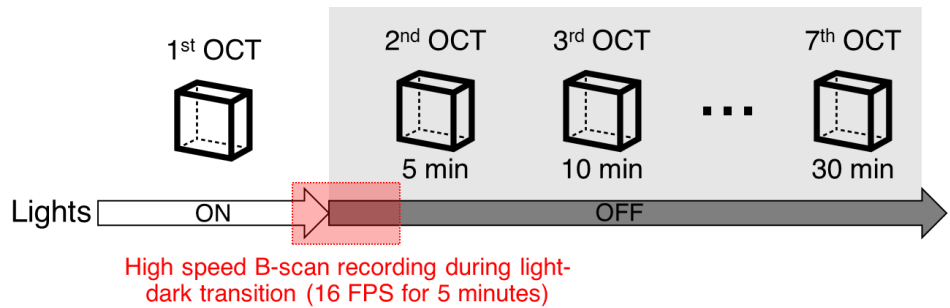

**Figure 7.** Image acquisition flow chart. The first OCT volume was acquired under ambient light conditions. After lights off, the second to seventh OCT volumes were sequentially acquired every 5 min up to 30 min. In between the first and second OCT volume acquisition, high-speed B-scan recording was performed to monitor retinal activity during a light-dark transitional moment.

**Image Analysis**

For quantitative analysis, retinal flattening was first implemented by realigning all A-lines in each OCT volume[61]. Next, adjacent 150 B-scans at the central region of OCT volume were averaged, followed by averaging 200 central A-lines, returning a single representative A-line per volume. Based on this A-line profile, retinal thickness and relative intensity value were measured. Each intensity profile was normalized based on the ONL intensity before relative intensity measurement in seven different OCT volumes. For the high-speed recording analysis, every 16 B-scans were first averaged, followed by retinal flattening, which returned a total 300 B-scans, and each B-scan represents 1-sec recording. 200 central A-lines were averaged for intensity and displacement measurement. To enhance the sampling density, each A-line was four times upsampled using linear interpolation. Image reconstruction and processing were done by MATLAB R2016a (MathWorks, Natick, MA).

**Statistical Analysis**

Data are expressed as mean ± standard deviation unless otherwise indicated. Statistical analysis was carried out with one-way repeated measures ANOVA to compare OCT measurements between seven different time points, followed by post hoc tests using Bonferroni correction to detect significant differences between the time points. The p-values less than 0.05 were considered significant in all tests. For high-speed recording analysis, Pearson's correlation analysis was conducted to examine the strength of a linear relationship between OCT measurements at different time points. All statistical analyses were performed with Origin 2020b (OriginLab, Northampton, MA).

**Data availability:** Data supporting the findings of this manuscript are available from the corresponding author upon reasonable request.

**Acknowledgments**: This research was supported in part by the National Institutes of Health (NIH) Grants R01EY023522, R01EY030101, R01EY030842, R01EY029673, and P30EY001792; by the Chicago Biomedical Consortium with support from the Searle Funds at the Chicago Community Trust; by the Richard and Loan Hill endowment; and by an unrestricted grant from Research to Prevent Blindness.

**Author contributions:** T.K. contributed to data collection, data processing, data analysis, and manuscript preparation; J.D. contributed to data analysis. X.Y. supervised the project and contributed to the study design, data analysis, and manuscript preparation.

**Competing interests:** The authors declare no competing interests.

**Corresponding authors:** Correspondence to Xincheng Yao.